\documentclass[preprint,preprintnumbers,amsmath,amssymb]{revtex4}
\usepackage[colorlinks=true, pdfstartview=FitV, linkcolor=blue, citecolor=red, urlcolor=magenta]{hyperref}
\usepackage{amssymb}
\usepackage{latexsym}
\usepackage{epsfig}
\usepackage{color}
\usepackage{amsmath}
\usepackage[utf8]{inputenc}
\allowdisplaybreaks
\begin{document}
\title{\textbf{Analysis of accretion disk around the Euler-Heisenberg Anti-de Sitter Black Hole }}
\author{ G.
Abbas $^{1,2}$ \footnote{ghulamabbas@iub.edu.pk }, H. Rehman $^1$
\footnote{hamzarehman244@gmail.com}}
\address{${}^1$ National Astronomical Observatories, Chinese Academy of Sciences, Beijing 100101, China\\
$^2$ Department of Mathematics, The Islamia University of Bahawalpur,\\ Bahawalpur Pakistan}

\date{}

\begin{abstract}
We demonstrate the investigation of thin accretion disc surrounding the Einstein- Euler-Heisenberg-Anti-de Sitter black hole. Additionally, we analyze the black hole event horizons and compute their effective potential and equations of motion. The specific energy, specific angular momentum, and specific angular velocity of the particles that move in circular orbits above the thin accretion disk are obtained. Also, the effects of parameters of the Einstein- Euler-Heisenberg-Anti de sitter black hole on specific angular velocity, specific heat energy, and specific angular momentum, have been discussed in detail. We display the positions of the innermost stable circular orbits and illustrate the effective potential. Furthermore, the circular orbits of black hole are obtained numerically and locates the position of the innermost stable circular orbit and event horizon.  Also, we study the Gamma ray burst emitted from the Einstein Euler-Heisenberg Anti de sitter black hole.


\end{abstract}
\maketitle
\section{Introduction}
Numerous astronomical studies have confirmed that the cosmos is expanding at an accelerated rate and General Relativity (GR) helps us to understand this phenomenon. Einstein's theory of GR has a lot of interesting predictions, but one of the most surprising is the existence of BH.
Black holes are the most fascinating objects in our universe and also the most mysterious one. The BH is a region in space where the force of gravity is so strong that not even light (the fastest known entity) in our universe can escape from it. When something crosses the event horizon, it collapses into the BH singularity, and the laws of physics no longer apply. Normally, BH are invisible to the naked eye, but they can be detected in a binary system or at the center of a galaxy using several approaches. The most realistic method of detecting BH is their influence on nearby substances, which is known as accretion. Over the last four decades, the fundamental perspectives of spherical accretion onto BHs have been extensively studied in \cite{Mm1}.
In 1934, Born and Infeld presented nonlinear electrodynamics(NLED) to make sure that the self-energy of a point-like charge is finite \cite{Mc4,Mc5}. Unless the 1980s, it was discovered that the reliable action for an open string ending on D-branes could be expressed precisely in nonlinear form \cite{Mc6}-\cite{Mc8}. The Born-Infeld BH is studied in \cite{Mc9}, as are the Einstein equations associated with Born-Infeld NLED.
Euler and Heisenberg suggested a novel method to interpret the electromagnetic field based on Dirac's positron theory \cite{Me2}, forecasting for the one-loop corrections to quantum electrodynamics (QED), and investigating the vacuum polarization in QED. This novel approach enables the NLED models to interpret the expansion of the cosmos at its beginning \cite{Me3}. In \cite{Me4} derived the Euler-Heisenberg (EH) BH solution by examining the one-loop lagrangian density associated with the Einstein field equation. Xiao-Xiong Zeng in \cite{Me5} looks into how the accretion flow models and QED affect the optical appearance of EH BH. Thermodynamics of the EH-AdS BH in \cite{Me6}

In astrophysics, accretion is the accumulation of particles into a central object by their gravitational attraction. It should be noted that among the most realistic scenarios for understanding the tremendously bright active galactic nuclei (AGNs) and the quasar is the accretion of matter onto the BHs. In \cite{Mm2} Bondi published an important work more than $40$ years ago. It was the first research on spherical accretion onto compact objects. In GR, Michl \cite{Mm3} analyzes the steady-state spherically symmetric flow of gas onto the Schwarzschild BH. He demonstrated that the accretion onto the BH must be transonic. Instead of the polytropic, equations of state have been used to study spherical accretion and winds in the framework of GR. Accretion has been examined in a variety of scenarios, including Reissner-N$\ddot{o}$rdstrom (RN) BH \cite{Mm6}, string cloud background \cite{Mm7} and charged BH \cite{Mm8} and Kerr-Newman BH  \cite{Mm9}- \cite{Mm11}. We are now interested in the study of the accretion disk around Euler and Heisenberg because one of the conceivable approaches for demonstrating the difference among GR and alternative theories is the study of accretion disks surrounding cosmic body.

The accretion disk is a flattened, circular or elliptical structure of energetic material that is orbiting the massive object such as a star. It can occur around protostar, BHs, neutron stars and white dwarfs. The mass of a BH can be increased by accretion disks, which means surrounding matter falls onto the disk \cite{Mm12}. Page \cite{Mm13}, analyzed the accretion disk mass through the rotating BH. The relativistic and radiation characteristics of thin accretion disks have been discussed in detail by Thorne \cite{Mm15}. The physical characteristics of the matter that forms thin accretion disks have been analyzed in spherically symmetric wormhole-space-time \cite{Mm16}. The characteristics of electromagnetic radiation released by the Kerr BH were examined in \cite{Mm17}. The optical properties of thin accretion disk surrounding a cosmic object are investigated using Einstein-Gauss-Bonnet gravity \cite{Mm18}. The theory of  non-relativistic accretion was extended by using the equatorial approximation to the axisymmetric and stationary space along with time of rotating BH \cite{Mm19}. The Schwarzschild BH accretion disk image was investigated by Luminet \cite{Mm20}. Numerous fixed values of the parameters were used to produce the image of the Kerr BH with the accretion disk \cite{Ma1,Ma2}. The further investigation of the accretion disk were discussed in Refs. \cite{Ma3,Ma8}. The enormous energy released in gamma ray bursts (GRBs), which are explosions of around $1\%$ of the solar mass, lasts just a few seconds \cite{Ma9,Ma10}. Most experts agree that the central engine is an accretion disk surrounding a newly generated BH. A pair of tremendously relativistic velocity jets powered by the BH accretion disk system formed the GRB if the jet is oriented in the direction of the spectator. The highly efficient gravitational energy of the progenitor star is needed to power such a large amount of energy over certain timescales, along with the relativistic jet. Furthermore, it is possible to verify the parameters of the EEH theory using the successful launch of two GRB jets.

The objective of this article is to examine the accretion disk around the EEH BH. The following is the structure of this article. We give a brief summary of the solutions EEH BH. Also, we examine the horizon of the BH in section II. We determine the effective potential and equation of motion in portion III. All the components of the thin accretion disk surrounding the EEH BH are investigated in section IV. In Section V, we describe the numerical analysis that involves the identification of the innermost stable circular orbits (ISCO). Furthermore, we explain the emission GRB from the BH accretion disk. Finally, we draw a conclusion in Section VI.
\section{Space-time}
This section provides a brief summary of the EH theory in conjunction with gravity. In GR, the $4D$ action with $\Lambda$ (cosmological constant) associated to NLED \cite{M27,M28} have the following form of action
\begin{equation}
S=\frac{1}{4\pi}\int_{M^{4}}d^{4}x\sqrt{-g}\Big(\frac{1}{4}(R-2\Lambda)-\mathcal{L}(F,G)\Big),\label{z1}
\end{equation}
where, $R$ stands for the Ricci scalar, $g$ denotes the metric determinant and $\mathcal{L}$ is the Lagrangian that relay on the electromagnetic invariants $F = \frac{1}{4}F_{ab}F^{ab}$ and $G=\frac{1}{4}F_{ab}\, ^*F^{ab}$, where $F_{ab}$ represents the the electromagnetic field strength and its dual is $\, ^*F^{ab}=\frac{\epsilon_{abcd}F^{cd}}{2\sqrt{-g}}$, the antisymmetric tensor $\epsilon_{abcd}$ fulfils $\epsilon_{abcd}\epsilon^{abcd}=-4$. The Lagrangian density for the Euler-Heisenberg NLED is \cite{Me2}
\begin{equation}
\mathcal{L}(F,G)= -F+\frac{a}{2}F^{2}+\frac{7a}{8}G^{2},\label{z2}
\end{equation}
where $a=\frac{8\alpha^{2}}{45m^{4}}$, $m$ is mass of electron and $\alpha$ is fine structure constant, which is EH parameter and we get the Maxwell electrodynamics $\mathcal{L}(F)=-F$ for $a=0$. In this perspective the electromagnetic field tensor $F^{ab}$, the possible frameworks for NLED are $F$ and $P$. Furthermore, there is $P$ framework that has a tensor $P_{ab}$ defined by
\begin{equation}
 P_{ab}=-\mathcal{L}_{X}F_{ab}+\, ^*F_{ab}\mathcal{L}_{G},\label{z3}
\end{equation}
hear $\mathcal{L}_{X}=\frac{dL}{dX}$. For the EH theory, $P$ has following form
\begin{equation}
 P_{ab}=(1-aF)F_{ab}-\, ^*F_{ab}\frac{7a}{4}G.\label{z4}
\end{equation}
In the EH NLED, the magnetic field $H$ and electric induction $D$ are both represented by this tensor, and the fundamental connections among these two fields (as well as the magnetic intensity $B$ and electric field $E$) are given by Eq. (\ref{a8}). The $P$ and $O$ are two different invariants that are related to the $P$ framework defined by
\begin{equation}
 P=\frac{-1}{4}P_{ab}P^{ab},\,\,\,\,   O=\frac{-1}{4}P_{ab}\, ^*P^{ab},\label{z5}
\end{equation}
with $\, ^*P^{ab}=\frac{\epsilon_{abcd}P^{cd}}{2\sqrt{-g}}$. The Hamiltonian is defined by the Legendre transformation of $\mathcal{L}$ as follows
\begin{equation}
\mathcal{H}(P,O)=\frac{-1}{2}F_{ab}P^{ab}-\mathcal{L},\label{z6}
\end{equation}
by omitting second and higher order terms in $a$, the Hamiltonian for the EH theory takes the form \cite{Mm25}
\begin{equation}
\mathcal{H}(P,O)=P-\frac{2}{2}P^{2}-\frac{7a}{8}O^{2}.\label{z8}
\end{equation}
In Ref. \cite{Mm24} the field equations are given by
\begin{equation}
\nabla_{a} P^{ab}=0, \,\,\,\, G_{ab}+\Lambda g_{ab}=8\pi T_{ab}.\label{z9}
\end{equation}
In the framework $P$, the energy momentum tensor for EH theory is defined by
\begin{equation}
T_{ab}= \frac{1}{4\pi}\Big((1-ap)P^{\alpha}P_{b\alpha}+g_{ab}(P-\frac{2}{2}P^{2}-\frac{7a}{8}O^{2})\Big).
\end{equation}
We find the solution of Eq. (\ref{z8}) for a static space-time, which is
\begin{equation}
ds^{2}=-f(r)dt^{2}+f^{-1}(r)dr^{2}+r^{2}(d\theta^{2}+sin^{2}\theta d\phi^{2}),\label{a1}
\end{equation}
with $f(r)=1-\frac{2m(r)}{r}$. The symmetry of space-time permits the non-vanishing components of the electromagnetic field, which is restricted to an electric charge of $Q$, so, we have
\begin{equation}
P_{ab}=\frac{Q}{r^{2}}\delta^{0}_{[a}\delta^{0}_{b]},\label{a2}
\end{equation}
hence, the electromagnetic invariants are
\begin{equation}
P_{ab}=\frac{Q^{2}}{2r^{4}}, \,\,\,\,\,\,\ O=0,\label{a3}
\end{equation}
inserting Eq.(\ref{a3}) into the $(t,t)$ component of equation Eq.(\ref{z9}), we get
\begin{equation}
\frac{dm}{dr}=\frac{Q^{2}}{2r^{2}}-\frac{aQ^{4}}{8r^{6}}+\frac{\Lambda r^{2}}{2},\label{a4}
\end{equation}
the metric function is derived by integrating the above equation, which has following form
\begin{equation}
f (r)=1-\frac{2 M}{r}+\frac{Q^2}{r^2}-\frac{\Lambda  r^2}{3}-\frac{a Q^4}{20 r^6},\label{a5}
\end{equation}
where, $Q$ and $M$ stand for the BH charge and mass, $a$ denotes the EEH parameter, and $\Lambda$ represents the cosmological constant, which is either negative or positive. In Eq. (\ref{a5}), the condition for which $a = 0$, coincides with the RN-AdS BH solution. The solution of $RN$ with electromagnetic Lagrangian $\mathcal{L}(F)=-F$ for the coupled Maxwell electromagnetic and gravitational field. The system behaves more like a Schwarzschild rather than the $RN-\Lambda$. It should be noted that $r = 0$ is the singular point and is much stronger than $RN-\Lambda$ with an opposite sign \cite{Mm22,Mm23}.
To examine the accretion disk around the EEH-AdS BH, we determine the BH horizons by assuming $f(r) = 0$, so
\begin{equation}
f (r)=1-\frac{2 M}{r}+\frac{Q^2}{r^2}-\frac{\Lambda  r^2}{3}-\frac{a Q^4}{20 r^6}=0,\label{a6}.
\end{equation}
In \textbf{Fig. 1}, we can find how many event horizons would exist for the considered BH, by counting the singular points on every curve. This demonstrates whether the curve coincide with $r-axis$, and the number of intersections tells us how many zeros there are. The red curve correspond to the RN-AdS BH $(a=0)$ that has two horizons. The green and blue curves represent the three event horizons of the EEH-AdS BH, while the magenta, cyan and black curves have one event horizon.
\begin{figure}
\includegraphics[width=.4\linewidth, height=1.5in]{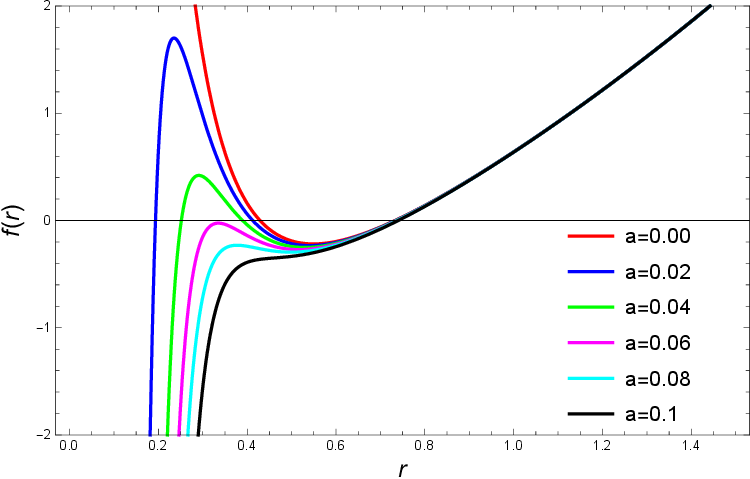}
\caption{The plot illustrated between $f (r)$ and $r$ is represented for $M=1$, $\Lambda=-3$, and $Q=0.8$. The red line correspond to the ($RN-\Lambda$) BH for $a=0$}.
\end{figure}

\section{Equation of motion}
Here, we determine the effective potential and equations of motion in order to determine the dynamic behavior of the system. Thus, we take into account the Lagrangian $\mathcal{L}$ for a particle in the vicinity of the EEH-AdS BH presented in Eq.(\ref{a1}). The Lagrangian $\mathcal{L}$ is given by
\begin{equation}
\mathcal{L}=\frac{1}{2} g_{\mu\nu}\frac{dx^{\mu}}{ds}\frac{dx^{\nu}}{ds}=\frac{1}{2}\epsilon,\label{a7}
\end{equation}
\begin{equation}
\mathcal{L}=\frac{1}{2} g_{tt}\Big(\frac{dt}{ds}\Big)^{2}+ g_{rr}\Big(\frac{dr}{ds}\Big)^{2}+ g_{\theta\theta}\Big(\frac{d\theta}{ds}\Big)^{2}+ g_{\phi\phi}\Big(\frac{d\phi}{ds}\Big)^{2}=\frac{1}{2}\epsilon.\label{a8}
\end{equation}
It is worthy to mention that the massive particle is represented by the value $ \epsilon = 1$, while $ \epsilon = 0$ is the value of the photon. The conserved angular momentum $L$ and energy $E$ for equatorial plane is
\begin{equation}
E=g_{tt}\Big(\frac{dt}{ds}\Big)=(1-\frac{2 M}{r}+\frac{Q^2}{r^2}-\frac{\Lambda r^2}{3}-\frac{a Q^4}{20 r^6})\frac{dt}{ds},\label{a9}
\end{equation}
\begin{equation}
L=g_{\phi\phi}\frac{d\phi}{ds}=r^{2}\frac{d\phi}{ds}.\label{c1}
\end{equation}
For massive particles, the geodesic equation of motion is given by
\begin{equation}
 \Big(\frac{dr}{ds}\Big)^{2}=E^{2}-(1-\frac{2 M}{r}+\frac{Q^2}{r^2}-\frac{\Lambda r^2}{3}-\frac{a Q^4}{20 r^6})(1+\frac{L^{2}}{r^{2}}),\label{c2}
\end{equation}
\begin{equation}
 \Big(\frac{dr}{d\phi}\Big)^{2}=\frac{r^{4}}{L^{2}}\Big(E^{2}-\Big(1-\frac{2 M}{r}+\frac{Q^2}{r^2}-\frac{\Lambda r^2}{3}-\frac{a Q^4}{20 r^6}\Big)\Big(1+\frac{L^{2}}{r^{2}}\Big)\Big),\label{c3}
\end{equation}
\begin{equation}
 \Big(\frac{dr}{dt}\Big)^{2}=\frac{1}{E^{2}}\Big(1-\frac{2 M}{r}+\frac{Q^2}{r^2}-\frac{\Lambda r^2}{3}-\frac{a Q^4}{20 r^6}\Big)^{2}\Big(E^{2}-\Big(1-\frac{2 M}{r}+\frac{Q^2}{r^2}-\frac{\Lambda r^2}{3}-\frac{a Q^4}{20 r^6}\Big)\Big(1+\frac{L^{2}}{r^{2}}\Big)\Big).\label{c4}
\end{equation}
It must noted be that the comprehensive dynamics of the system may be described by using Eqs. (\ref{c2}-\ref{c4}). In addition, we may calculate the effective potential using Eq. (\ref{c2}) as follows
\begin{equation}
V_{eff}=\Big(1-\frac{2 M}{r}+\frac{Q^2}{r^2}-\frac{\Lambda r^2}{3}-\frac{a Q^4}{20 r^6}\Big)\Big(1+\frac{L^{2}}{r^{2}}\Big).\label{b1}
\end{equation}
When we take into account Eq. (\ref{b3}) and get
\begin{equation}
V_{eff}=\Big(1-\frac{2}{\tilde{r}}+\frac{\tilde{Q^{2}}}{\tilde{r}^{2}}-\tilde{\Lambda} \tilde{r}^{2}-\frac{\tilde{a}\tilde{Q}^{4}}{20 \tilde{r}^{6}}\Big)\Big(1+\frac{\tilde{L}^{2}}{\tilde{r}^{2}}\Big),\label{b2}
\end{equation}
where
\begin{equation}
\tilde{r}=\frac{r}{M},\,\,\,\,\,\, \tilde{Q}=\frac{Q}{M},\,\,\,\,\,\tilde{L}= \frac{L}{M},
\tilde{a}=\frac{a}{M^{2}},\,\,\,\,\,\,\tilde{\Lambda}=\frac{\Lambda M^{2}}{3}, \label{b3}
\end{equation}
these are dimensionless variables.
\section{Accretion disk around EEH-AdS black hole}
In the present section, we will be able to discuss the physical characteristics of the accretion disk surrounding the EEH-AdS BH and determine all of its parameters. As we know, the particles move in a circular orbit above the disk, so we can compute the specific angular momentum $L$, specific energy $E$, specific angular velocity $\Omega$, and flux of the radiant energy $F$. All the quantities may be determined for every possible case. It should be noted that the physical characteristics of the accretion disk obey the particular structural equations that are associated with the conservation of mass, energy and angular momentum. These equations may be determined by the conservation of energy, angular momentum and mass. It is obvious that radius of the orbit affects the kinematic values, and we can get these by using the general formulas. We use dimensionless variables from Eq. (\ref{b3}) to evaluate all of the parameters \cite{M31,M30} as follows
\begin{equation}
 \Omega=\sqrt{\frac{1}{\tilde{r}^{3}}-\tilde{\Lambda}-\frac{\tilde{Q}^{2}}{\tilde{r}^{4}}+\frac{3 \tilde{a} \tilde{Q}^{4}}{20 \tilde{r}^{8}}},
\end{equation}
\begin{equation}
\tilde{E}=\frac{\Big(1-\frac{2}{\tilde{r}}+\frac{\tilde{Q^{2}}}{\tilde{r}^{2}}-\tilde{\Lambda} \tilde{r}^{2}-\frac{\tilde{a}\tilde{Q}^{4}}{20 \tilde{r}^{6}}\Big)}{\sqrt{1-\frac{3}{\tilde{r}}+\frac{2 \tilde{Q}^{2}}{\tilde{r}^{2}}-\frac{4 \tilde{a}\tilde{Q}^{4}}{20 \tilde{r}^{6}}}},
\end{equation}

\begin{equation}
\tilde{L}=\frac{\tilde{r}^{2}\sqrt{\frac{1}{\tilde{r}^{3}}-\tilde{\Lambda}-\frac{\tilde{Q}^{2}}{\tilde{r}^{4}}+\frac{3 \tilde{a} \tilde{Q}^{4}}{20 \tilde{r}^{8}}}}{\sqrt{1-\frac{3}{\tilde{r}}+\frac{2 \tilde{Q}^{2}}{\tilde{r}^{2}}-\frac{4 \tilde{a}\tilde{Q}^{4}}{20 \tilde{r}^{6}}}}.
\end{equation}
From Eq. (\ref{b1}), we get
\begin{multline}
\frac{d^{2}V_{eff}}{dr^{2}}=\frac{1}{40 \tilde{r}^8(5 \tilde{r}^4(2 \tilde{Q}^2+(\tilde{r}-3) \tilde{r})-\tilde{a} \tilde{Q}^4)^3}\Big(21 \tilde{a}^4 \tilde{Q}^{16}+5 \tilde{a}^3 \tilde{Q}^{12} \tilde{r}^4(\tilde{r}(8 \tilde{\Lambda} \tilde{r}^3-63 \tilde{r}+202)-144 \tilde{Q}^2)\\+25 \tilde{a}^2 \tilde{Q}^8 \tilde{r}^8(-6 \tilde{Q}^2 \tilde{r}(48 \tilde{\Lambda} \tilde{r}^3-63 \tilde{r}+176)+3 \tilde{r}^2 (\tilde{r}(2 \tilde{r}(4 \tilde{\Lambda} \tilde{r}(30 \tilde{\Lambda} \tilde{r}^3-29 \tilde{r}+38)+33)-239)+278)+364 \tilde{Q}^4)\\ -1000 \tilde{a} \tilde{Q}^4 \tilde{r}^{12}(-4 \tilde{Q}^4 \tilde{r} (3 \tilde{\Lambda} \tilde{r}^3-4 \tilde{r}+27)+\tilde{Q}^2 \tilde{r}^2(\tilde{r}(2\tilde{r}(\tilde{\Lambda} \tilde{r}(88 \tilde{\Lambda} \tilde{r}^3-39 \tilde{r}+28)-9)+39)+56)+\\3 \tilde{r}^3 (\tilde{r} (\tilde{r} (\tilde{r} (\tilde{\Lambda} (\tilde{r} (2 \tilde{r} (5 \tilde{\Lambda} (\tilde{r}-6) \tilde{r}+3)-1)+6)-7)+33)-51)+18)+36 \tilde{Q}^6)\\+10000 \tilde{r}^{16}(-2 \tilde{Q}^6 \tilde{r}(4 \tilde{\Lambda} \tilde{r}^3-9 \tilde{r}+32)+\tilde{Q}^4 \tilde{r}^2(\tilde{r}(2 \tilde{r}(6\tilde{\Lambda} \tilde{r}(5 \tilde{\Lambda} \tilde{r}^3-3 \tilde{r}+4)+5)-69)+126)+2 \tilde{Q}^2\tilde{ r}^3 \\(\tilde{r} (\tilde{r} (\tilde{\Lambda} \tilde{r} (\tilde{r} (\tilde{r} (\tilde{\Lambda} \tilde{r} (17 \tilde{r}-72)-6)+41)-36)-9)+41)-54)\\+\tilde{r}^4 (\tilde{r} (\tilde{r} (\tilde{r} (\tilde{\Lambda} (\tilde{r} (\tilde{r} (\tilde{r} (3 \tilde{\Lambda} (\tilde{r} (2 \tilde{r}-15)+30)-2)+18)-54)+36)-1)+12)-36)+36)+12 \tilde{Q}^8)\Big).
\end{multline}
We impose some restrictions on the circular motion of the particles, so we put $V_{eff} =0$, $\frac{dV_{eff}}{dr} =0$ and $\frac{d^{2}V_{eff}}{dr^{2}}=0$.
Now we are interested to calculate ISCO of EEH BH, which exists at a local minimum of effective potential, so by using $V_{eff} =0$, $\frac{dV_{eff}}{dr} =0$ and $\frac{d^{2}V_{eff}}{dr^{2}}=0$, we can determine ISCO. The analytical solution of $V_{eff} =0$, $\frac{dV_{eff}}{dr} =0$ and $\frac{d^{2}V_{eff}}{dr^{2}}=0$ is not possible, so we calculate ISCO numerically. Further, details are given in section \textbf{VI}.
It should be noted that the Stefan Boltzmann law \cite{M29} is used to determine the flux of radiation as follows
\begin{equation}
F(r)= \frac{-\dot{M}_{0}}{4\pi\sqrt{-g}}\frac{\Omega_{,r}}{(\tilde{E}-\Omega\tilde{L})^{2}} \int^{r}_{r_{isco}}(\tilde{E}-\Omega\tilde{L})\tilde{L_{,r}dr},
\end{equation}
where the mass accretion rate is denoted by $\dot{M_{0}}$. It is important to note that flux conservation laws can be used to get radiant energy above the disk between radial distances and the ISCO. \textbf{Figure 2} illustrates the pattern of the energy flux of a thin accretion disk around the EEH BH for dimensionless variables $\tilde{Q} = 0.8$, $\tilde{\Lambda }= 3$ and various values of $\tilde{a}$.
\begin{figure}
\includegraphics[width=.7\linewidth, height=2in]{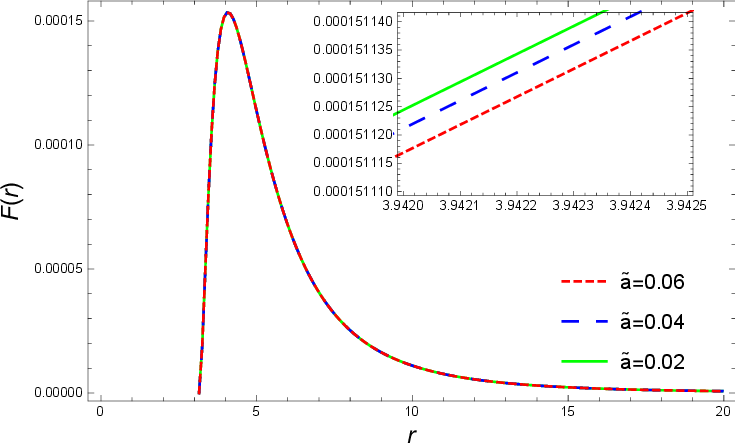}
\caption{ The \textit{F(r)} of an accretion disk surrounding the EEH BH for Q = 0.8 and $\dot{M}_{0}$ = 1}.
\end{figure}

\begin{figure}
(a)\includegraphics[width=.4\linewidth, height=1.5in]{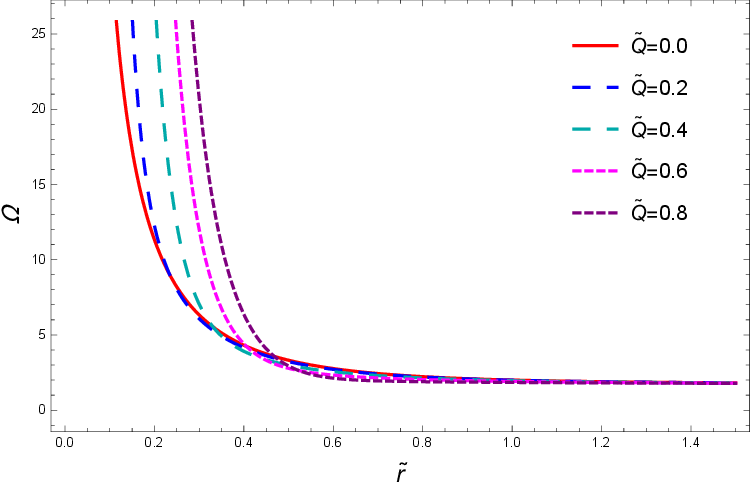}
(b)\includegraphics[width=.4\linewidth, height=1.5in]{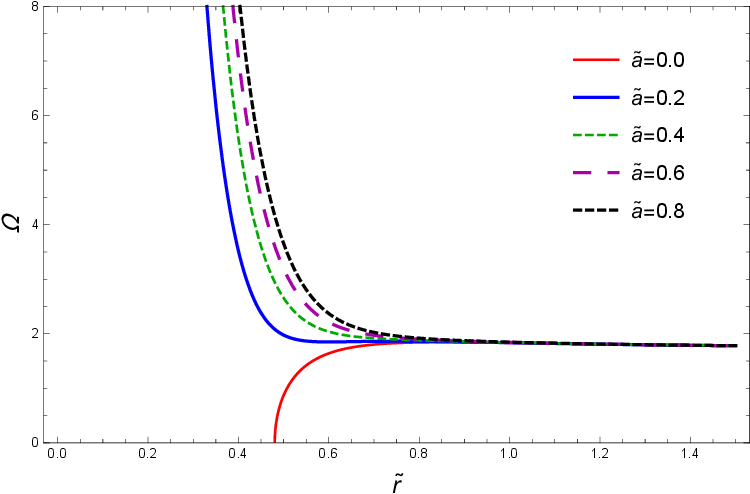}
(c)\includegraphics[width=.4\linewidth, height=1.5in]{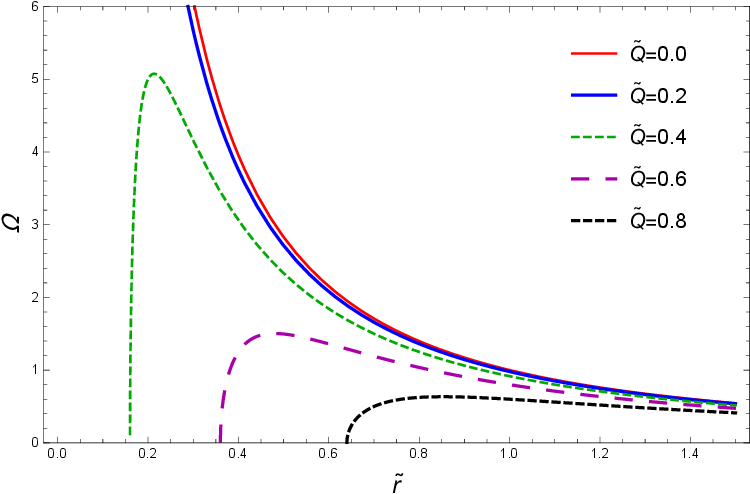}
(d)\includegraphics[width=.4\linewidth, height=1.5in]{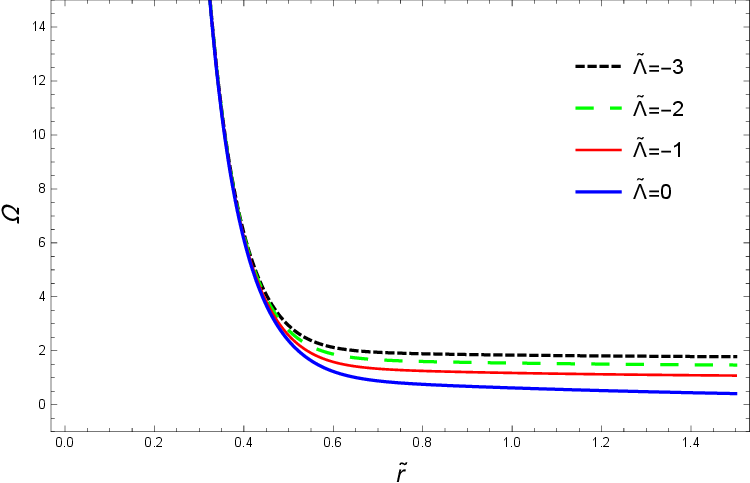}
\caption{The plot displays the angular velocity $\Omega$ verses $\tilde{r}$ of thin accretion disk and around the BH. (\textbf{a}) for $\tilde{a}=0.5$, $\tilde{\Lambda}=-3$ and for distinct values of $\tilde{Q}$. (\textbf{b}) for $\tilde{Q}=0.8$, $\tilde{\Lambda}=-3$ and for altered values of $\tilde{a}$. (\textbf{c}) for $\tilde{a}=0.5$, $\tilde{Q}=0.8$ and for various values of $\tilde{\Lambda}$. (\textbf{c}) for $\tilde{a}=0$, $\tilde{\Lambda}=0$ and for various values of $\tilde{Q}$ }
\end{figure}
 In \textbf{Fig. 3}, the plot is drawn between angular velocity and the dimensionless variable $\tilde{r}$ in terms of dimensionless quantities $\tilde{Q},~~~\tilde{a}$, and $\tilde{\Lambda}$.
In plots \textbf{3.a} and \textbf{3.b}, the angular velocity increases slightly as the value of the dimensionless variables $\tilde{Q}$ and $\tilde{a}$ increases, but in plot \textbf{3.c}, the angular velocity decreases as the value of the dimensionless variable $\tilde{\Lambda}$ increases, and in plot \textbf{3.d}, we recover the behavior of the angular velocity of the Schwarzschild BH by assuming $\tilde{Q}=0$ and $\tilde{a}=0$, while angular velocity of RN BH can be restored by considering $\tilde{a}=0$ with various values of $\tilde{Q}$.

\begin{figure}
(a)\includegraphics[width=.4\linewidth, height=1.5in]{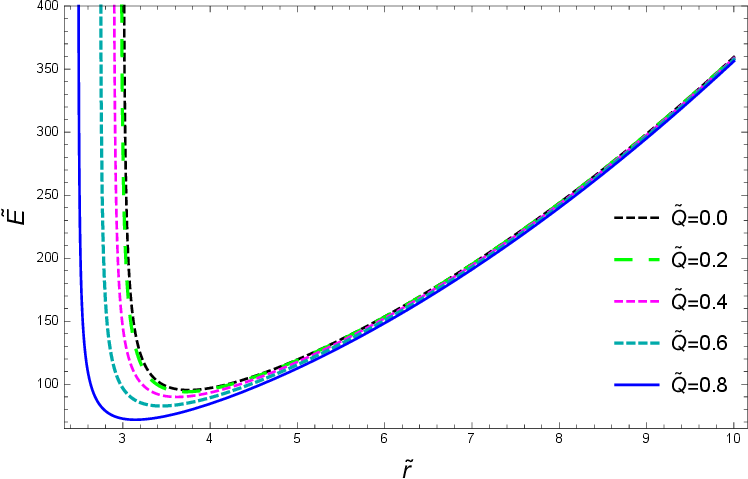}
(b)\includegraphics[width=.4\linewidth, height=1.5in]{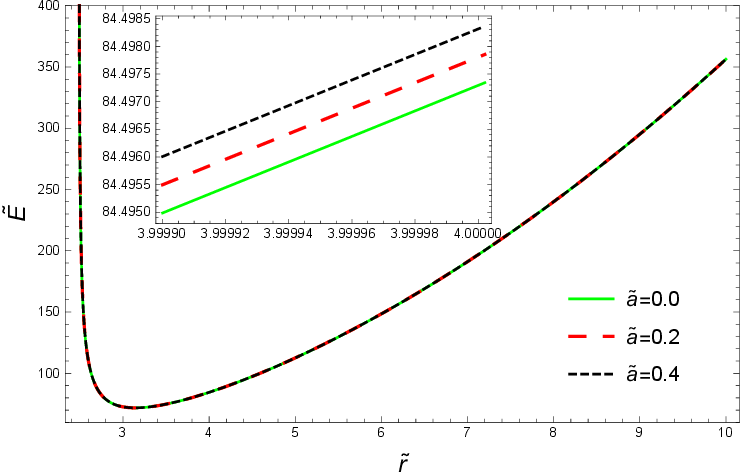}
(c)\includegraphics[width=.4\linewidth, height=1.5in]{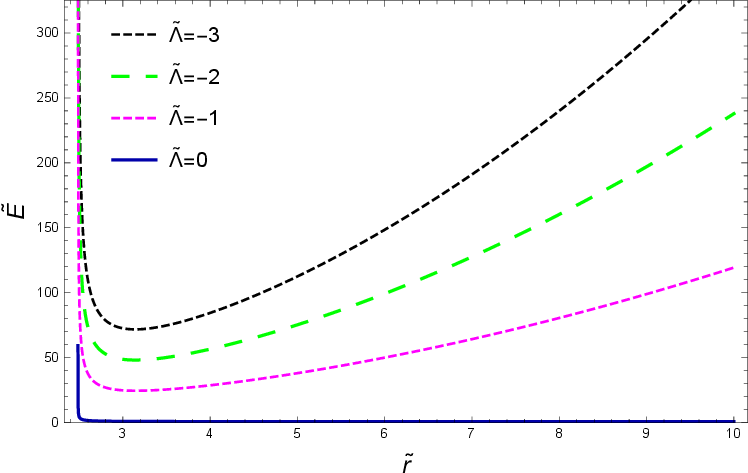}
(d)\includegraphics[width=.4\linewidth, height=1.5in]{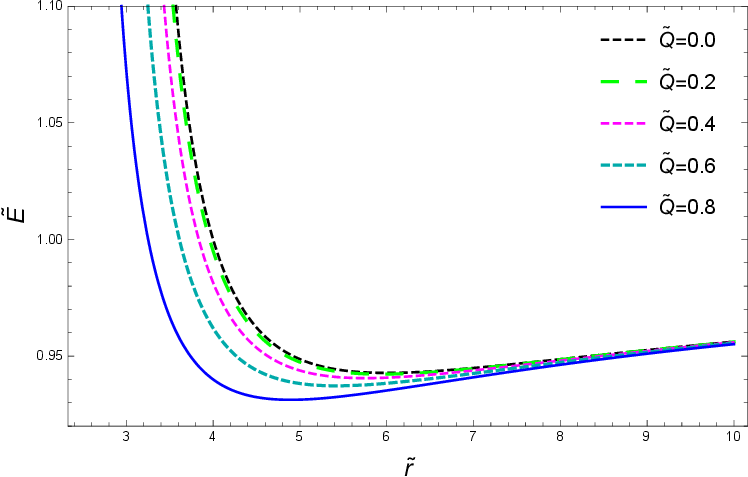}
\caption{The plot between $\tilde{E}$ verses $\tilde{r}$ around the BH. (\textbf{a}) for $\tilde{a}=0.5$, $\tilde{\Lambda}=-3$ and for various values of $\tilde{Q}$. (\textbf{b}) for $\tilde{Q}=0.8$, $\tilde{\Lambda}=-3$, and for different values of $\tilde{a}$. (\textbf{c}) For $\tilde{a}=0.5$, $\tilde{Q}=0.8$ and for distinct values of $\tilde{\Lambda}$. (\textbf{d}) for $\tilde{a}=0$, $\tilde{\Lambda}=0$ and for altered values of $\tilde{Q}$ }
\end{figure}

The specific energy $\tilde{E}$ verses $\tilde{r}$ for altered values of $\tilde{a}$, $\tilde{Q}$ and $\tilde{\Lambda}$ as represented in \textbf{Fig. 4}. From \textbf{Fig}. \textbf{4a} and \textbf{4b}, it is clear that in the thin accretion disk, the specific energy $\tilde{E}$ of the particle rapidly decreases and attains a minimum value and then it continuously increases. Also, from the illustration of \textbf{4c}, we can observe that the specific energy of the particle initially decreases to attain a minimum value for $\tilde{\Lambda} = -3$, $\tilde{\Lambda} = -2$, $\tilde{\Lambda} = -1$, and then increases gradually. Also, for $\tilde{\Lambda} = 0$, the energy of the particle decreases and become constant. It should be worth noticing that, by imposing some restrictions, i.e.,
$\tilde{a} = 0$,~~ $\tilde{\Lambda} = -3$ on the EEH BH, we retrieve the specific energy $\tilde{E}$ to $\tilde{r}$ of the thin accretion disk of Schwarzschild and RN BHs, as shown in illustration \textbf{4d}. In Plot\textbf{ 4d}, the black doted line represents the specific energy of thin accretion disk around Schwarzschild BH, while the remaining lines represent the specific energy of thin accretion disk around RN BH.

\begin{figure}
(a)\includegraphics[width=.4\linewidth, height=1.4in]{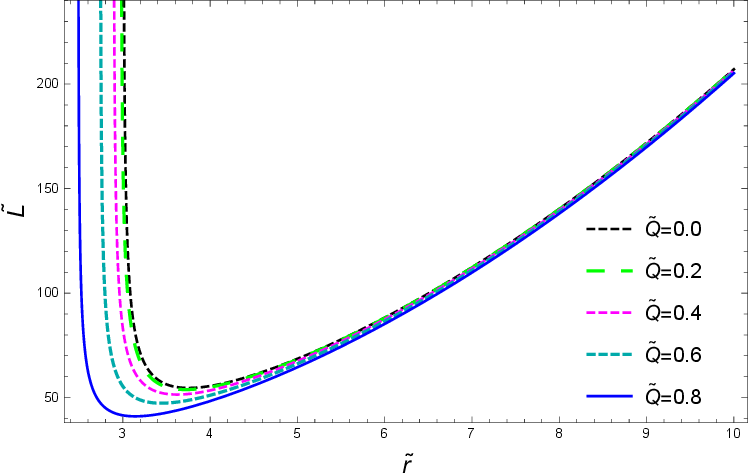}
(b)\includegraphics[width=.4\linewidth, height=1.5in]{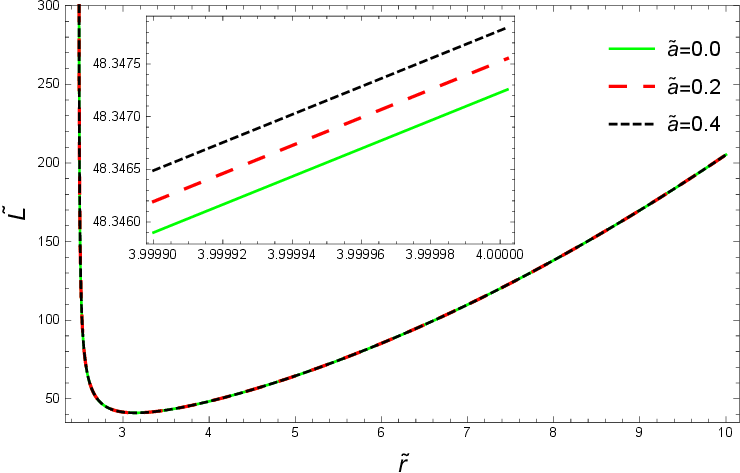}
(c)\includegraphics[width=.4\linewidth, height=1.5in]{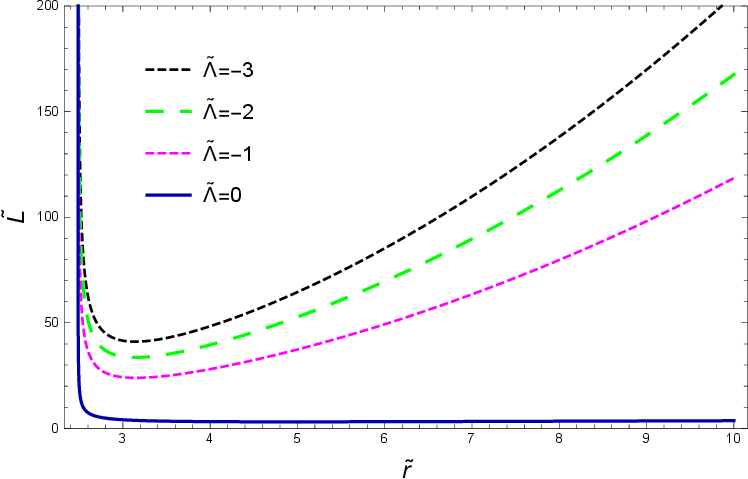}
(d)\includegraphics[width=.4\linewidth, height=1.5in]{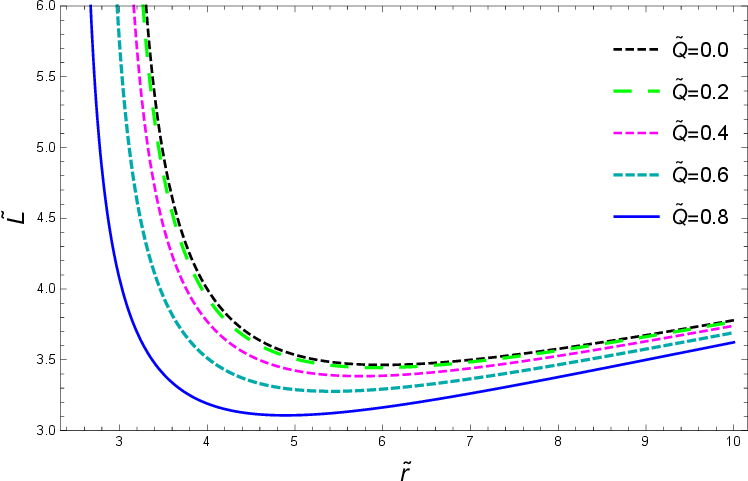}
\caption{The plot between $\tilde{L}$ verses $\tilde{r}$ of thin accretion disk around the BH. (\textbf{a}) For $\tilde{a}=0.5$, $\tilde{\Lambda}=-3$, and for different values of $\tilde{Q}$. (\textbf{b}) For $\tilde{Q}=0.8$, $\tilde{\Lambda}=-3$ and for various values of $\tilde{a}$. (\textbf{c}) For $\tilde{a}=0.5$, $\tilde{Q}=0.8$ and for distinct values of $\tilde{\Lambda}$. (\textbf{d}) For $\tilde{a}=0$, $\tilde{\Lambda}=0$ and for altered values of $\tilde{Q}$ }
\end{figure}
In Fig.\textbf{5}, we have represented the specific angular momentum $\tilde{L}$ in accordance with $\tilde{r}$ for several values of $\tilde{Q}$, $\tilde{a}$ and $\tilde{\Lambda}$.
From, illustrations \textbf{5a} and \textbf{5b}, we see that for distinct values of $\tilde{Q}$, $\tilde{a}$, the angular momentum $\tilde{L}$ decreases for a small value of $\tilde{r}$, which is around $\tilde{r}=3$, and then increases. In illustration \textbf{5c}, we see that for $\tilde{\Lambda}=-3,~~\tilde{\Lambda} = -2$, and $\tilde{\Lambda} = -1$, the angular momentum decreases for a small value of $\tilde{r}$ and then increases, but for $\tilde{\Lambda} = 0$, the angular momentum decreases to become constant. Furthermore, we also analyzed the behavior of the specific angular moment of the thin accretion disk around the Schwarzschild and RN BH from EEH BH by assuming $\tilde{a} = 0$, $\tilde{\Lambda} = 0$ for different values of $\tilde{Q}$, which is shown in \textbf{Fig. 5d}. In \textbf{Fig. 5d}, the black dot line represents the behavior of the specific angular momentum of the thin accretion disk of the Schwarzschild BH, while the remaining lines indicate the behavior of the specific angular momentum of the thin accretion disk around the RN BH.
\section{Numerical Analysis}
In this section, we investigate the ISCO, effective potentials and indicate the position of the ISCO of the EEH BH. In GR the smallest marginally stable orbit around the BH is said to be ISCO. It must be emphasized that the accretion disk ISCO play a vital role in the BH because it marks the inner edge of the BH \cite{Ma11}. It should be noted that for $a = 0$ , $Q = 0$, and $\Lambda = 0$ the EEH BH reduce to Schwarzschild BH with $r_{isco}=6M$. For $a = 0$, $\Lambda = 0$, we recover RN BH, and the position of ISCO for $0<Q<1$ is $(4,6)$. The location of horizons and ISCO is displayed in tables \textbf{I}, \textbf{II} and \textbf{III}.
\begin{table}[ht]
    \caption{For $ Q = 0.8$, $\Lambda = -3$, and different values of EEH BH parameter $a$, the position of the horizons and the ISCO are shown.} 
    \centering 
    \begin{tabular}{c c c c c c} 
        \hline 
          a \ & Inner horizon \ & Middle horizon \ & Outer horizon \ & ISCO\\ [0.7ex] 
        \hline 
       0.0 &-& 0.438 & 0.7328 & 3.14996\\ 
        0.02 &0.18& 0.42 & 0.73425 & 3.14998\\
        0.04 &0.225& 0.38 & 0.73515 & 3.14999\\
        0.06 &-& -& 0.73675 &(0.3241, 0.3470, 3.15001)\\
        0.08 &-& -& 0.732 &(0.3331, 0.4203, 3.15003)\\
        0.1 &-& -& 0.741 &(0.3499, 0.4481, 3.15004)\\
      \hline 
    \end{tabular}
    \label{table:nonlin}
\end{table}

\begin{table}[ht]
    \caption{The positions of the horizons and the ISCO for $a = 0.02$, $Q = 0.8$, and various values of $\Lambda$ are given.} 
    \centering 
    \begin{tabular}{c c c c c c} 
        \hline 
         $\Lambda$ \ & Inner horizon \ & Middle horizon \ & Outer horizon \ & ISCO\\ [0.7ex] 
        \hline 
      -3 &0.193&0.41 & 0.721 & 3.14998\\ 
        -2 &0.1939& .402 & .84 & 3.15572\\
        -1 &0.19391& 0.392 & 1.02& 3.17239\\
        0 &0.194&0.384& 1.596& 4.89078\\
        1 &0.195&0.37& - & (0.910447,3.10018)\\
        2 &0.194&0.36& - & (0.739501,3.11973)\\
      \hline 
    \end{tabular}
    \label{table:nonlin}
\end{table}

\begin{table}[ht]
    \caption{For $a = 0.005$, and $\Lambda = -3$, the position of the horizons and the ISCO for various values of $Q$ are represented around the EEH BH.} 
    \centering 
    \begin{tabular}{c c c c c c} 
        \hline 
         $Q$ \ & Inner horizon \ & Middle horizon \ & Outer horizon \ & ISCO\\ [0.7ex] 
        \hline 
       0.0 &-& - & 0.998 & 3.76053\\ 
        0.2 &-& - & 0.98 & 3.72831\\
        0.4 &-& - & 0.95 & (0.102336,0.144972,3.62808)\\
        0.6 &0.128&0.182& 0.88 & 3.44678\\
         0.8 &0.126&0.42& 0.72 & 3.14996\\
          1.0 &0.1351&-& - & (0.593385,1.14954,2.60655)\\
      \hline 
    \end{tabular}
    \label{table:nonlin}
\end{table}

\begin{figure}
\includegraphics[width=.6\linewidth, height=2in]{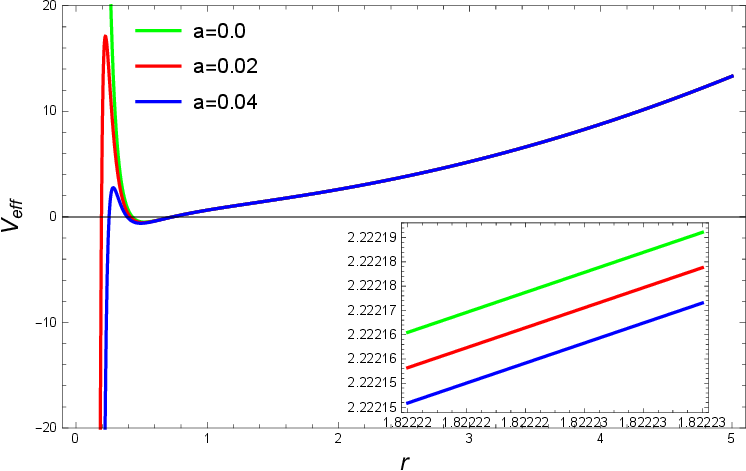}
\caption{The behavior of effective potential for $L=1$, $Q$=0.8, $\Lambda$=-3 and altered values of $a$.}
\end{figure}
The profile of effective potential versus $r$ is shown in \textbf{Fig. 6} for $L = 1$, $Q = 0.8$, $\Lambda$ = -3, and various values of the EEH BH parameter $a$. From graph it is clear that as the value of $a$ goes up, its effective potential goes down. Also, by considering $a = 0$, we recover the behavior of the effective potential for RN-AdS BH, which is indicated by the green curve.

\begin{figure}
a) \includegraphics[width=.3\linewidth, height=1.3in]{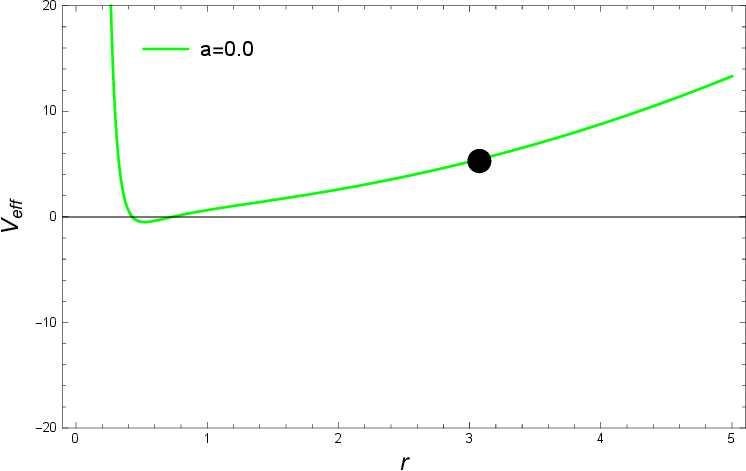}
b) \includegraphics[width=.3\linewidth, height=1.3in]{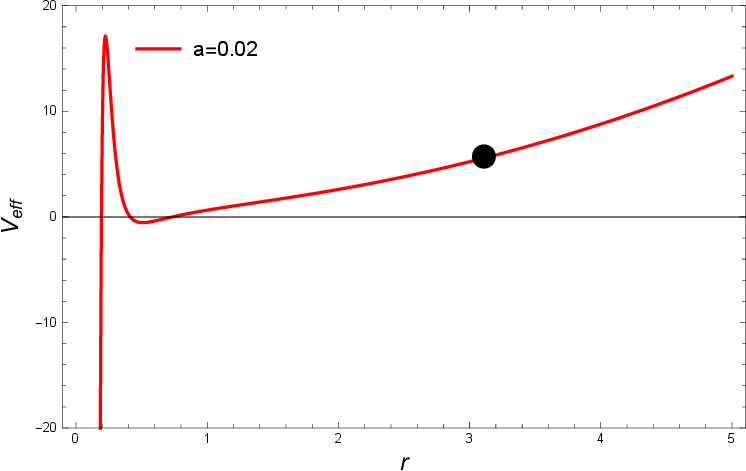}
c) \includegraphics[width=.3\linewidth, height=1.3in]{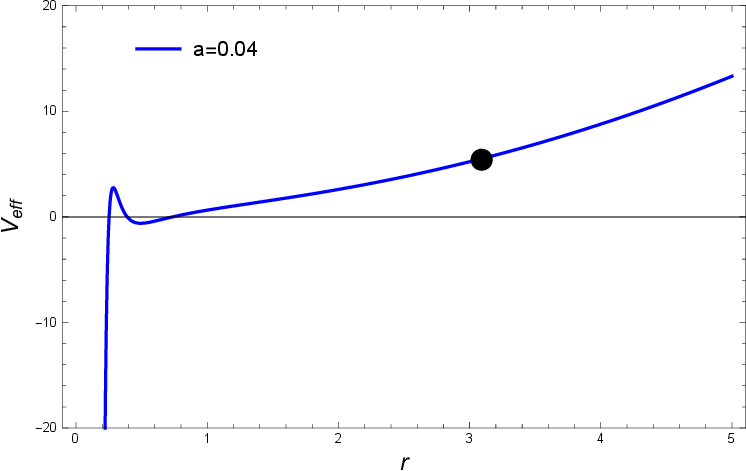}
\caption{The position of ISCO represented by dots for $L=1$, $Q$=0.8, $\Lambda$=-3. (a) For $a=0.0$, ISCO is located outside the outer event horizon at $r_{isco}=3.1496$. (b) For $a=0.02$, ISCO is at $r_{isco} =3.1498$, which is outside the outer event horizon. (c) The ISCO for $a=0.04$, we have $r_{isco}=3.1499$, which lies far from the outer event horizon.}
\end{figure}

In addition, we displayed the position of ISCO in \textbf{Fig. 7} by considering $L = 1$, $Q = 0.8$ and $\Lambda$ = -3 for different values of $a$. In \textbf{Fig. 7} \textbf{a}, \textbf{b}, and \textbf{c}, the ISCOs with $r_{isco} = 3.1496$, $r_{isco} = 3.1498$ and $r_{isco} = 3.1499$, respectively, are located outside the outer event horizons.

\begin{figure}
\includegraphics[width=.6\linewidth, height=2.1in]{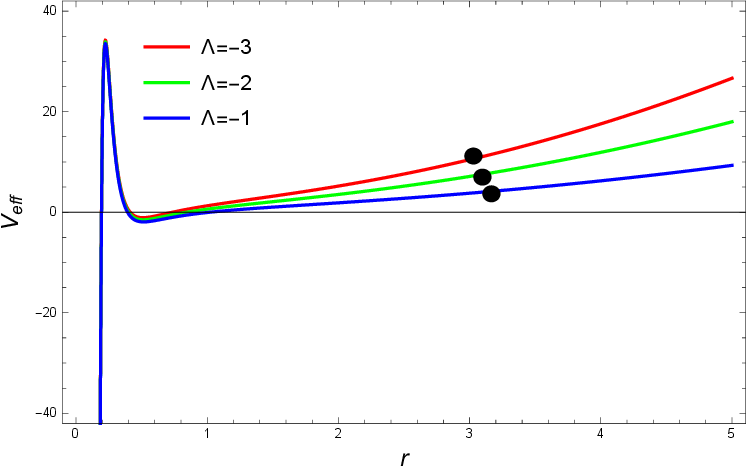}
\caption{ Depicts the profile of effective potential for $L=1$, $a$=0.02, $Q$=0.8, with different values of $\Lambda$ also the location of ISCO indicated by solid circles.}
\end{figure}
Figure 8 shows the effective potential corresponding to $r$ for $L= 1$, $a = 0.02$, $Q = 0.8$ and various values of $\Lambda$. The plot clearly shows that as the value of $\Lambda$ grows, its effective potential drops. ISCO is located at $r_{isco} = 3.14998$ for $\Lambda = -3$, $r_{isco} = 3.15572$ for $\Lambda = -2$, and $r_{isco} = 3.17239$ for $\Lambda = -1$. We also locate the ISCO position, which is indicated by dark circles.
\begin{figure}
\includegraphics[width=.6\linewidth, height=2in]{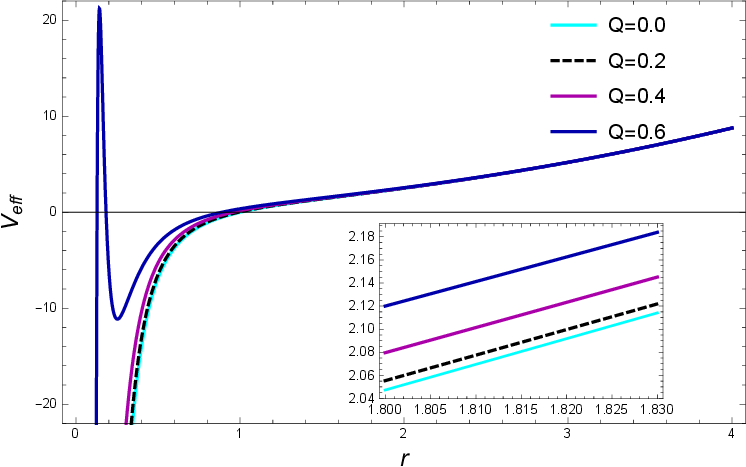}
\caption{The profile of effective potential for $L=1$, $a$=0.005, $\Lambda$=-3 and various values of $Q$.}
\end{figure}
\textbf{Figure 9} depicts the profile of effective potential vs. $r$ for $L = 1$, $a = 0.005$, $\Lambda = -3$ and several $Q$ values. From the plot, it is clear that as the value of $Q$ increases, so does its effective potential. Furthermore, by assuming $Q = 0$, we recover the behavior of the effective potential for the Schwarzschild-AdS BH, as shown by the cyan curve.

\begin{figure}
\includegraphics[width=.4\linewidth, height=1.5in]{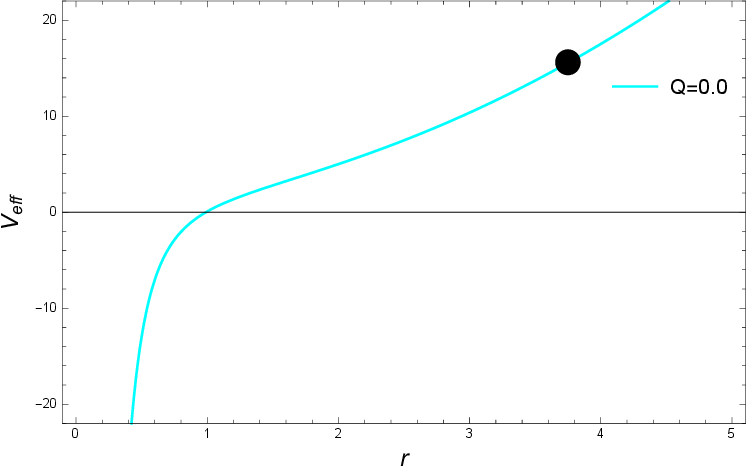}
\includegraphics[width=.4\linewidth, height=1.5in]{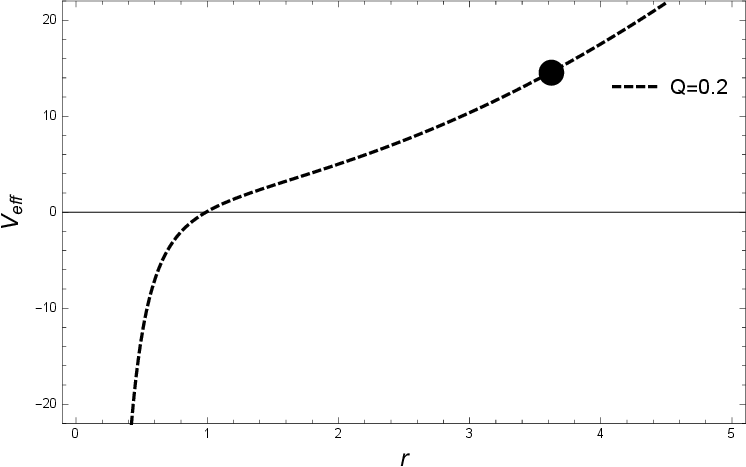}
\includegraphics[width=.4\linewidth, height=1.5in]{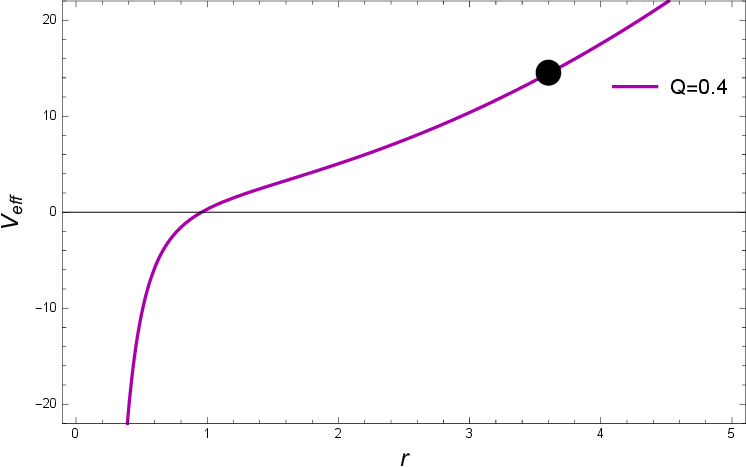}
\includegraphics[width=.4\linewidth, height=1.5in]{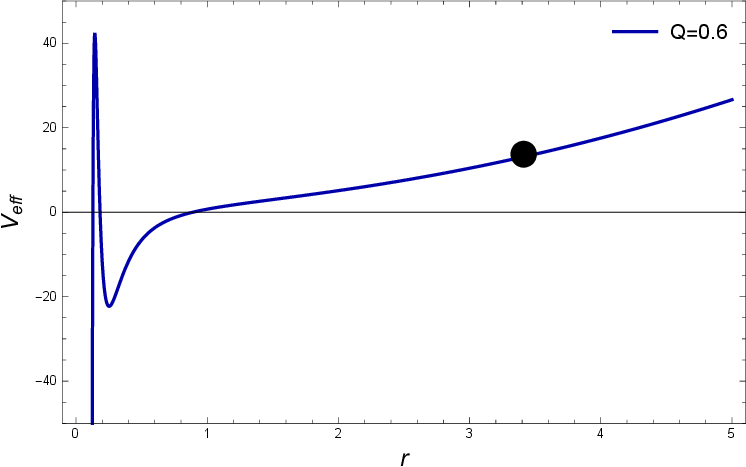}
\caption{The solid circle denotes the location of ISCO for $L=1$, $a$=0.005, $\Lambda$=-3, and altered values of charge Q. (a) ISCO is located outside the outer horizon at $r_{isco} = 3.76053$ for $Q=0$.(b) For $Q=0.2$, the ISCO is $r_{isco}=3.72831$, which is outside the outer horizon. (c) The ISCO for $Q=0.4$ is $r_{isco} =3.62808$, which is located beyond the outer event horizon. (d) ISCO is located outside the outer event horizon at $r_{isco} =3.44678$ for $Q=0.6$.}
\end{figure}
Furthermore, we locate the position of ISCO in \textbf{Fig. 10} by assuming $L = 1$, $a = 0.005$, $\Lambda = -3$ and numerous values of $Q$.

\subsection{Gamma ray Burst}
In the given portion, we investigate the GRB emitted from the BH. We should speculate regarding GRBs from the BH accretion disk. It is important to note that the the outer event horizon is usually smaller than outer stable circular orbit. By assuming a certain efficiency $\eta$ for transferring gravitational energy to gamma-ray energy, we can constrain the parameters of the EEH BH. The ordinary mass of the progenitor is $M\sim10M_{\odot}$, the total energy of an usual gamma-ray burst is $\sim10^{52}$ erg, and the outer stable circular orbit $x$ which are given in Ref.\cite{Mc4}.
\begin{equation}
E_{\gamma}=\eta\frac{M}{x}. \label{s1}
\end{equation}
Consider $ \eta<0.1$, which gives
\begin{equation}
x<\eta\frac{M}{E_{\gamma}}\sim18\eta_{-1}M_{1}E_{\gamma,52}^{-1}, \label{s2}
\end{equation}
where
\begin{equation}
Q=10^{k}\times Q_{k},\,\,\,\ M_{1}=\frac{M}{10M_{\bigodot}}. \label{s3}
\end{equation}
Here $M_{1}$ denotes the unit of solar mass. 
\section{Conclusions}
Accretion discs around supermassive BHs are the main source of gravitational information in systems with intense gravity, and they also show how space and time are arranged around them. Researchers in \cite{Mc1}-\cite{Mc3} have examined the luminosity of accretion disks in the space with time of a static BH occupied by dark matter and of BHs surrounded by dark matter with isotropic or anisotropic pressure, respectively.
In the present study, we have investigated the thin accretion disk around static and spherically symmetric EEH BH. Furthermore, we have examined the event horizons, the effective potential and equation of motion of the EEH BH. We have determined the specific angular velocity, specific energy, and specific angular momentum of particles that traverse in a circular orbit around the EEH BH. Also, we have demonstrated the the flux of radiant energy above the disk.
Moreover, the energy flux of thin accretion around the EEH BH for various values of $\tilde{a}$ are shown in \textbf{Fig. 2}
In addition to this, we have investigated the variations in the specific energy, the specific angular momentum, ISCO of the particles orbiting around the BH, and the angular velocity as a result of the various values of the parameters of the EEH BH, which are shown in \textbf{Figs}. \textbf{3}, \textbf{4}, and \textbf{5}. We have plotted the effective potential and location of ISCO is represented in \textbf{Fig} \textbf{5}. Furthermore, we have investigated the location of ISCO and the event horizon is given in table \textbf{III}. Also, we recover the Schwarzschild BH with $\tilde{r}_{isco}= 6M$ and RN BH with $\tilde{r}_{isco} = (4,6)$ by imposing some conditions on EEH. Finally, we analyzed the GRB with more realistic efficiency $\eta$, obtained from simulations and specific extreme events.
\section*{Acknowledgements}
The work of G. Abbas has been partially supported by the National Natural Science Foundation of China under project No. 11988101.
He is grateful to compact object and diffused medium Research Group at NAOC led by Prof. JinLin Han for excellent hospitality and friendly environment.
G. Abbas is also thankful to The Islamia University of Bahawalpur, Pakistan for the grant of 06 months the study leave.

\end{document}